\documentclass[sigconf, language=french,
language=german, language=spanish, language=english,anonymous=false]
{acmart}
\renewcommand\footnotetextcopyrightpermission[1]{}

\AtBeginDocument{%
  \providecommand\BibTeX{{%
    \normalfont B\kern-0.5em{\scshape i\kern-0.25em b}\kern-0.8em\TeX}}}
\acmConference[]{}{}{} 

\begin{document}

\title{Incorporating Spatial Awareness in Data-Driven Gesture Generation for Virtual Agents}

\author{Anna Deichler}
\email{deichler@kth.se}
\orcid{1234-5678-9012}
\authornotemark[1]
\affiliation{%
  \institution{KTH Royal Institute of Technology}
  \streetaddress{}
  \city{Stockholm}
  \state{}
  \country{Sweden}
  \postcode{}
}

\author{Simon Alexanderson}
\email{simonal@kth.se}
\affiliation{%
  \institution{KTH Royal Institute of Technology}
  \city{Stockholm}
  \country{Sweden}
}

\author{Jonas Beskow}
\email{beskow@kth.se}
\orcid{0000-0003-1399-6604}
\affiliation{%
  \institution{KTH Royal Institute of Technology}
  \city{Stockholm}
  \country{Sweden}
}


\begin{abstract}
This paper focuses on enhancing human-agent communication by integrating spatial context into virtual agents' non-verbal behaviors, specifically gestures. Recent advances in co-speech gesture generation have primarily utilized data-driven methods, which create natural motion but limit the scope of gestures to those performed in a void. Our work aims to extend these methods by enabling generative models to incorporate scene information into speech-driven gesture synthesis. We introduce a novel synthetic gesture dataset tailored for this purpose. This development represents a critical step toward creating embodied conversational agents that interact more naturally with their environment and users.

\end{abstract}






\maketitle


    
\section{Introduction}

Human communication is rooted in a spatial and physical context. In situated communicative settings people engage in referential communication in identifying, describing, or giving instructions related to objects, locations, or people. They make use of multimodal expressions, consisting of spatial language and non-verbal behaviors, such as gaze and pointing gestures. For a virtual agent to function effectively in a spatially contextual environment, it must be capable of interpreting and responding to both verbal and environmental cues. This involves understanding objects, locations, and other contextual factors that influence how gestures should be formed and directed. For instance, pointing at an object while describing it, or making eye contact with an interlocutor during a conversation, requires an integration of scene understanding with gesture generation.
Non-verbal behavior generation in virtual agents has a long history.  Early work in gesture synthesis for virtual agents was rule-based, based on inverse kinematics and procedural animation to generate co-speech gestures \cite{cassell2001beat,kopp2004synthesizing,marsella2013virtual}. Rule-based approaches, however, often require laborious manual tuning and often struggle to achieve a natural motion quality. In recent years data-driven solutions have become primary and supervised learning systems for co-speech gesture generation and have achieved high naturalness in certain settings. Most of these systems take audio as input \cite{alexanderson2020style,hasegawa2018evaluation,ferstl2020adversarial}, and are limited to generating beat gestures aligned with speech. Some recent work extends audio conditioning with text conditioning to obtain semantic information \cite{kucherenko2020gesticulator,deichler2023diffusion}.
On the one hand, this made the generation of highly natural motion possible, but on the other hand, it also narrowed the scope of these behaviors.  In particular, there has been a surge in co-speech gesture generation research due to novel co-speech gesture datasets consisting of speech and motion modalities. These co-speech generation systems focus on generating gestures that lack spatial information. However, for any kind of situated interaction, these agents need exposure to the spatial context. In our work, we aim to develop generative models that extend existing speech-driven gesture synthesis frameworks by incorporating scene information. In this paper, we propose a novel synthetic gesture dataset that augments an existing co-speech dataset with multimodal referring expressions, as a first step towards extending speech-driven gesture synthesis with scene information conditioning.

\section{Related work}
\subsection{Situated Gesture generation}

Generation of situated gestures, such as deictic expressions, has been considered in virtual agents for a long time \cite{noma2000design,rickel1999animated,lester1999deictic,kappagantula2019automatic}. Similar approaches have been employed in referential human-robot communication \cite{fang2015embodied, sauppe2014robot, rubies2023remote}. All of these systems are based on rules and/or procedural generation, which limits the degree of naturalness, fluidity, and variability that can be achieved in the resulting animations, and makes it difficult to match the gesture quality produced by contemporary generative models for co-speech gestures as described above. Recent work started considering the data-driven generation of pointing gestures \cite{deichler2023learning}.

\subsection{Scene conditioning in human motion generation}
In recent years there has been more focus on scene-conditioned human motion generation \cite{wang2024move,wang2022humanise,huang2023diffusion}. These works in human-scene interaction focus mainly on non-conversational human locomotion generation, ranging from generating static human poses in 3D environments \cite{zhang2020generating} to temporally extended human motion synthesis, conditioning on the 3D layout of the environment and human-object interaction \cite{zhang2024force}. Conditional motion generation works include diffusion-based \cite{huang2023diffusion} and reinforcement learning-based methods. 

Currently gesture generation in virtual agents and scene-conditioned human motion generation are separate fields, but in order to have situated gesturing in virtual agents there should be more interaction between these fields.

\section{Dataset generation}
In order to train conversational embodied agents that are situated in a spatial context, datasets are needed that contain the relevant environment information. Currently, there is a lack of such datasets. One way to acquire such a dataset is to extend and combine existing datasets that contain the necessary annotations in themselves. In recent years, generating synthetic data has offered a solution to machine learning problems, where data is scarce and difficult to access. It has become common practice in the domain of text generation\cite{liu2024best}, and there are also works in motion generation \cite{black2023bedlam},\cite{mehta2024fake}. Similar strategies could be applied for generating synthetic datasets that could help extend the current capabilities of co-speech gesture generation agents.

In the realm of deictic expressions, which involve pointing and other indicating gestures, there is a critical need for systems that not only recognize speech and text but also incorporate the physical environment. This integration allows virtual agents to perform gestures that are not only relevant to the spoken content but are also spatially aware, enhancing their utility in real-world interactions. Such advancements are vital for developing more sophisticated and interactive virtual agents that can operate effectively within dynamic environments.

As a step towards exposing current co-speech gesture agents to a spatial context, we combined an existing co-speech gesture dataset, which mainly contains beat gestures, with an existing pointing gesture dataset. The co-speech gesture dataset we used\cite{ferstl2021expressgesture} is one of the largest natural conversation datasets of synchronized motion capture and speech recordings with a duration of approximately 6 hours. The dataset contains recordings of a single English speaker. Synthesizing human-like pointing gestures in virtual avatars requires an appropriate full-body motion capture dataset with diverse and accurate target locations covering the 3D space surrounding the character. We utilized the solely available dataset that includes both motion capture data of pointing gestures and the corresponding 3D target locations \cite{deichler2022towards}. This dataset also includes recorded speech as the actor
is pointing. Example utterances in single target pointing are simple demonstrative expressions, such as "This one!", "I want that one!", "Give me that one over there!". The speech data is synchronized with mocap.  

\begin{figure}[h]
  \centering
  \includegraphics[height=5.5cm]{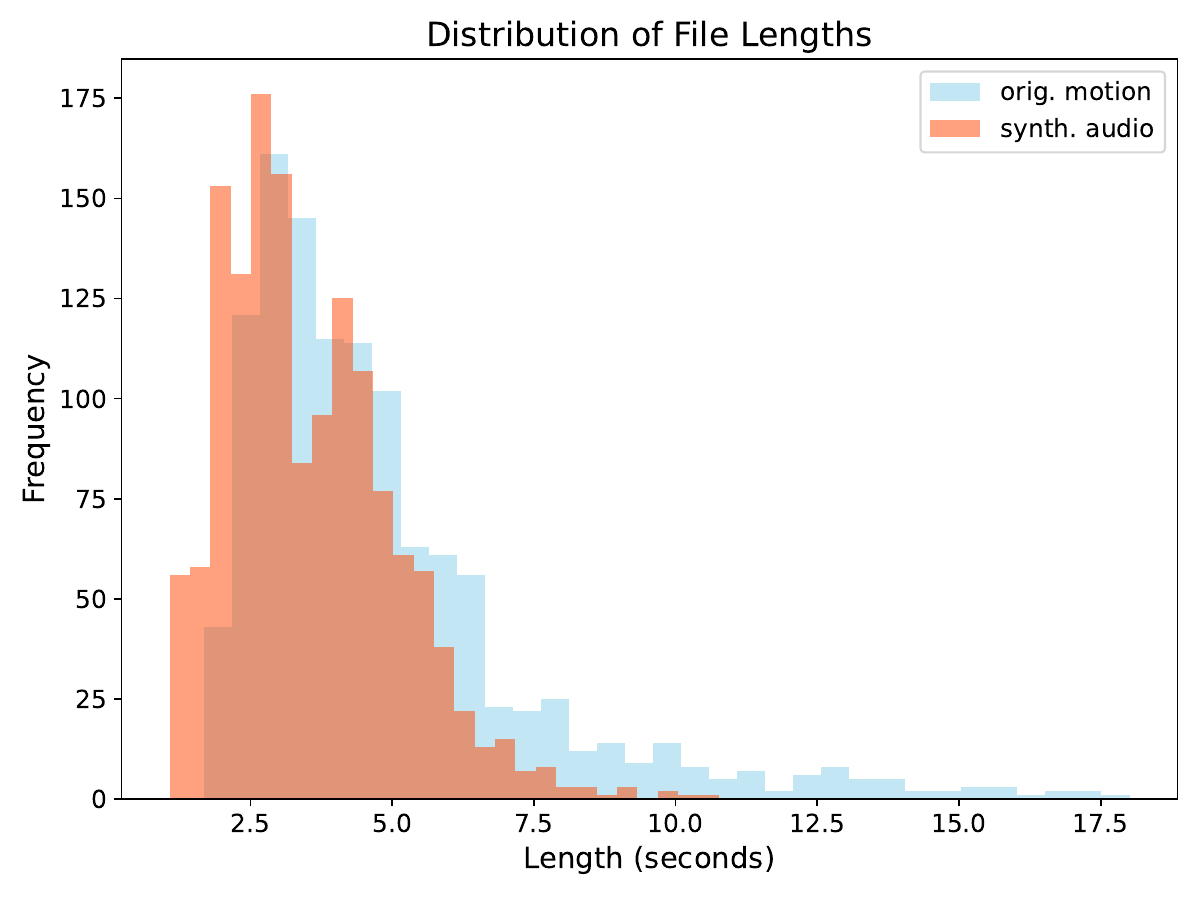}
  \caption{Distribution of length of pointing motion clips and synthesized audio clips.}
  \label{fig:clip_lengths}
  \vspace{-1\baselineskip}
\end{figure}

We replaced these simple demonstratives with synthesized ones to better match the other co-speech gesture dataset's acoustic and semantic qualities.
\paragraph{Extending the pointing gesture dataset}
The original pointing dataset contained approximately 140 isolated single target pointing gestures, ranging from 0.5 to 9 seconds in duration. To account for the small size of this dataset, all motion files were mirrored and each of the individual pointing gesture clips was stretched in time in the range of [0.7, 1.6] three times. This resulted in a pointing gesture dataset of 1160 clips. The distribution of pointing gesture lengths can be seen on Figure \ref{fig:clip_lengths}. The mean duration of the pointing gesture was 4.85 seconds.
\paragraph{Text generation}
 We used GPT-4  \cite{achiam2023gpt} to generate simple demonstrative references for the pointing gesture dataset. LLMs using Generative Pre-trained transformers (GPT) can be utilized to generate large amounts of task-specific text with prompts appropriate for the given task. To have semantic similarity between the beat gesture and pointing gesture datasets, we provided the transcripts of the beat gesture dataset in prompting the LLM as contextual information.  Furthermore, each utterance should be an exophoric referring expression that could be coupled with a pointing gesture. The main instruction prompt can be found in Appendix \ref{app:prompts}. 

\begin{figure}[h]
  \centering
  \includegraphics[width=.9\columnwidth]{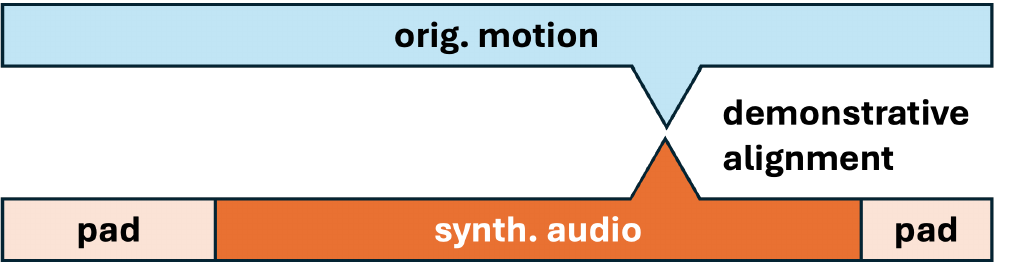}
  \caption{Alignment of demonstratives in motion and speech, with pre-and post-padding of audio clips.}
  \label{fig:alignment}
  \vspace{-1\baselineskip}
\end{figure}

\paragraph{Speech synthesis}
Based on the generated text, speech segments were synthesized using a commercially available TTS engine \cite{11labs}. Emphasis on demonstrative words in the synthesized utterances text was added using the Speech Synthesis Markup Language (SSML) phoneme tags. In total,  1400 sentences were synthesized.
\begin{quote}\texttt{
<speak><emphasis level=strong>THIS</emphasis> one, towering above the rest, is the oldest tree in the forest.</speak>}
\end{quote}
\begin{quote}\texttt{
 <speak>On this shelf, can you spot <emphasis level='strong'>
 THAT</emphasis> vase there? It's handmade.</speak>
 }
 \end{quote}

 \begin{figure}[h]
  \centering
  \includegraphics[height=5.5cm]{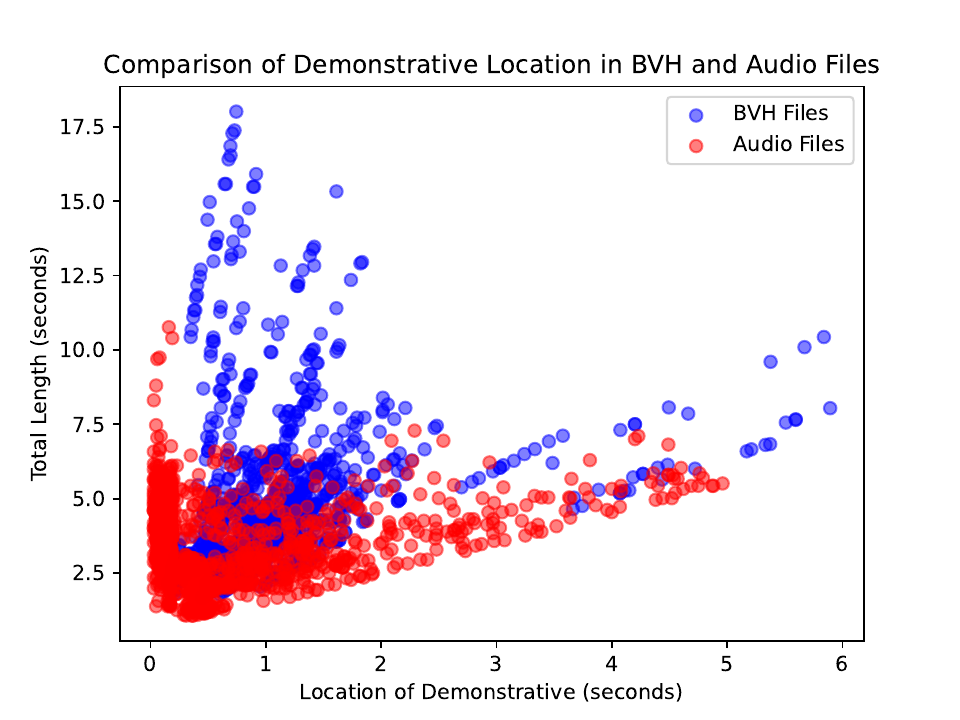}
  \caption{Demonstrative location against total length (in seconds) for motion (BVH) and audio files.}
  \label{fig:timing_comp}
  \vspace{-1\baselineskip}
\end{figure}

\paragraph{Matching the generated speech segments with pointing gestures}
The generated speech segments were matched with the pointing gestures using the Hungarian algorithm. 

Figure \ref{fig:clip_lengths} compares the length of the synthesized audio clips against the length of the pointing gestures.

The cost function considered the difference in length between synthetic and pointing segments, as well as the different in alignment between demonstratives, i.e. penalizing matches where the synthetic event does not temporally align with the original demonstrative event (e.g., starting before or ending after the demonstrative). The aim was to maintain consistency with the original data's temporal distribution of demonstrative within the utterance as seen on Figure \ref{fig:timing_comp}. 

As can be seen from Figures \ref{fig:clip_lengths} and \ref{fig:timing_comp}, there is a discrepancy between animation and audio lengths. This is intentional since we wanted the speech segments to be fully contained within the animation clips after the demonstratives are aligned in the two streams. The speech segments were subsequently padded with leading and trailing silence, to the same length as the animations, see Figure \ref{fig:alignment}. 

\begin{figure}[h]
  \centering
  \includegraphics[height=3.5cm]{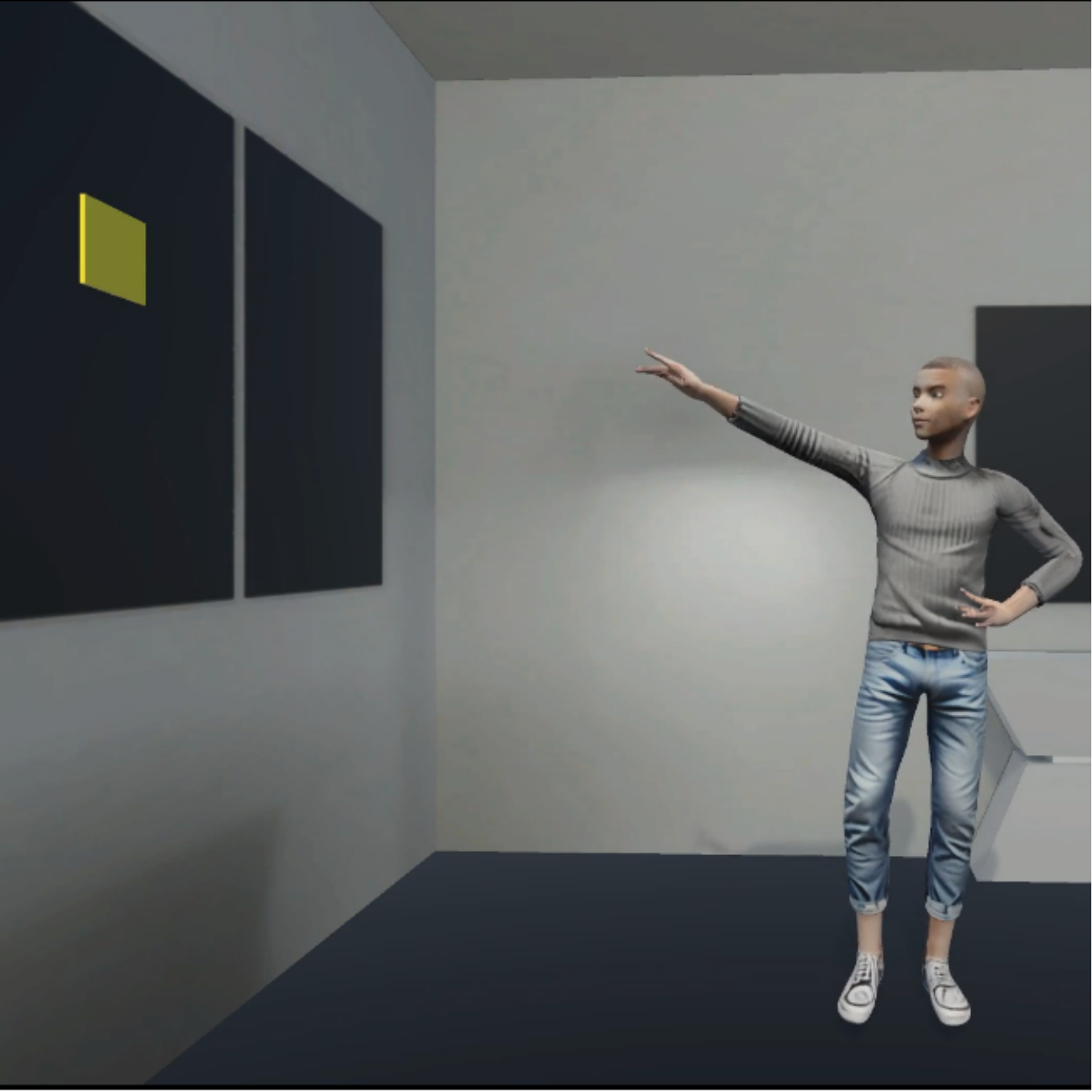} 
  \hspace{5mm} 
  \includegraphics[height=3.5cm]{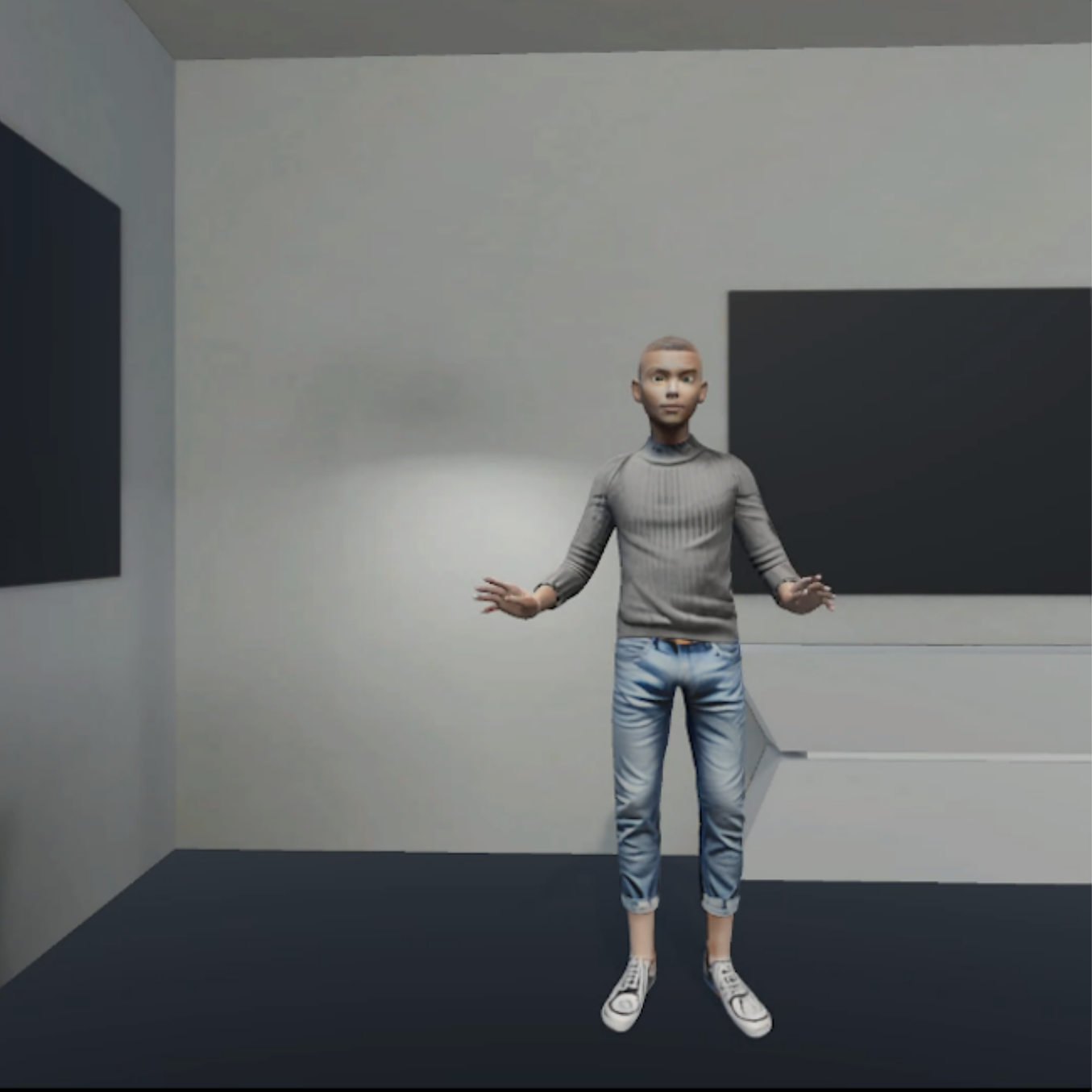} 
  \caption{Visualization of dataset examples containing synchronized clips of audio and gesture for (a) pointing and (b) beat gestures.}
  \label{fig:agent}
\end{figure}

Figure \ref{fig:agent} shows a visualization of a pointing gesture and a beat gesture from the dataset. Further examples can be found on the demo website. 
\footnote{https://huggingface.co/spaces/annadeichler/spatial-gesture}.




\section{Conclusions and future work}
In this paper, we highlighted one of the major shortcomings of current data-driven gesture generation systems that the gestures are generated completely in isolation, without any spatial context that would make the gestures situated. We provided a synthetic dataset, which could serve as a benchmark to test computational models that are capable of joint co-speech and referential gesture generation. These simple, synthetic datasets are beneficial to extend the capabilities of current co-speech gesture generation models, but in the future, more datasets are needed that have contextual information, for example, spatial context. These datasets could include annotations about the scene, for example, details about the objects and spatial configurations present in the environment during each recorded interaction. Processing this enriched dataset will require sophisticated data handling strategies to ensure that the model can generalize from training data to real-world applications.

Another important consideration is how to properly evaluate situated gestures in a virtual agent. We propose to build such evaluation methods on previous gesture generation evaluations. For example, referential games can be utilized for pointing accuracy, as in \cite{deichler2023learning}. For gesture naturalness and appropriateness, we can refer to the methodologies used in the GENEA challenge \cite{kucherenko2023genea}. 

Going forward, we believe that generating synthetic datasets or relying on multiple separate datasets in training will be necessary to achieve agents capable naturalistic of situated gesture and this dataset provides a first benchmark to develop methods that integrate spatial awareness into the natural gesture generation capabilities of virtual agents.
\bibliographystyle{ACM-Reference-Format}
\bibliography{sample-base}

\appendix
\section{GPT-4 prompts}
\label{app:prompts}
The demonstrative references were generated using this prompt:
\begin{quote}
    Demonstratives in exophoric reference are used to indicate concrete physical entities in space. For example, I was thinking of sitting down. See that bench there, we could sit there.
Here 'that' would be the demonstrative.
I want you to generate examples like the following, which connect to a story.
The rule is that each generated sentence should contain one exophoric reference.
Also, put the demonstrative in the middle of the sentence.
Examples:
" See that bench over THERE, we could sit there", 
"Look in the middle of the room, THAT small dog is my new best friend.", 
"We could get there faster. See THAT bike over there.", 
"Look at the top shelf. THAT is the vase I wanted to buy".
\end{quote}

After this, summaries of the \cite{ferstl2021expressgesture} dataset transcripts were provided as additional context. Furthermore, the location of the demonstrations was also directly prompted to cover better the original pointing dataset's demonstrative location in the accompanying pointing gesture motion sequence.










\end{document}